\documentclass[runningheads]{llncs}
\usepackage{graphicx}
\usepackage{booktabs}
\usepackage{cleveref}
\usepackage{amssymb}
\usepackage{todonotes}
\begin{document}
\title{Going Beyond the Cookie Theft Picture Test: Detecting Cognitive Impairments using Acoustic Features}
\titlerunning{Detecting Cognitive Impairments using Acoustic Features}
%
%
\author{Franziska Braun\inst{1}\and 
Andreas Erzigkeit\inst{2} \and
Hartmut Lehfeld\inst{3} \and
Thomas Hillemacher\inst{3} \and
Korbinian Riedhammer\inst{1} \and 
Sebastian P. Bayerl\inst{1}} 
\authorrunning{Braun et al.}
%
\institute{Technische Hochschule Nürnberg Georg Simon Ohm, Germany\and
Psychiatrische Klinik und Psychotherapie, Universitätsklinikum Erlangen, Germany\and
Klinik für Psychiatrie und Psychotherapie, Universitätsklinik der Paracelsus Medizinischen Privatuniversität, Klinikum Nürnberg, Germany
\email{franziska.braun@th-nuernberg.de}}
%
%
\maketitle              

\begin{abstract}
Standardized tests play a crucial role in the detection of cognitive impairment. 
Previous work demonstrated that automatic detection of cognitive impairment is possible using audio data from a standardized picture description task. 
The presented study goes beyond that, evaluating our methods on data taken from two standardized neuropsychological tests, namely the German SKT and a German version of the CERAD-NB, and a semi-structured clinical interview between a patient and a psychologist. 
For the tests, we focus on speech recordings of three sub-tests: reading numbers (SKT 3), interference (SKT 7), and verbal fluency (CERAD-NB 1).
We show that acoustic features from standardized tests can be used to reliably discriminate cognitively impaired individuals from non-impaired ones. 
Furthermore, we provide evidence that even features extracted from random speech samples of the interview can be a discriminator of cognitive impairment.
In our baseline experiments, we use OpenSMILE features and Support Vector Machine classifiers. 
In an improved setup, we show that using wav2vec 2.0 features instead, we can achieve an accuracy of up to 85\%.

\keywords{dementia screening \and pathological speech \and paralinguistics \and neuropsychological tests}
\end{abstract}
\section{Introduction}
In geriatric patients, dementia represents one of the most common condition seen in the psychiatric consultation service of a general hospital. 
According to the WHO, over 55 million people worldwide were living with dementia in 2020 \cite{WHO21}. 
This number will nearly double every 20 years, reaching 78 million in 2030 and 139 million in 2050. 
The estimated annual global cost of dementia currently exceeds US\$ 1.3 trillion and is expected to rise to US\$ 2.8 trillion by 2050, of which more than half is care costs.

Dementia is characterized by a loss or decline of function; in addition to memory impairments, patients exhibit one or more of aphasia, apraxia, agnosia or impairments of executive function.
These symptoms can relate to different neurological conditions (e.g., Alzheimer's).
Due to its typically insidious onset, dementia is in many cases detected too late. 
Early diagnostic clarification with the resulting possibility of a rapid start of treatment is key to slowing the progression of dementia and thus achieving a gain in quality of life for the patient and their family caregivers.
Dementia screening and monitoring enable early detection, classification and tracking of cognitive decline.

In addition to medical examinations (e.g., brain imaging), a combination of medical and psychological history taking, cognitive testing, and the use of rating scales is the gold standard for dementia screening in clinical or research settings \cite{cooper05}.
To that end, standardized tests play a key role in clinical routine since they aim at minimizing subjectivity by measuring performance on a variety of cognitive tasks.
Tests typically target both short- and long-term memory and cover tasks such as naming, memorizing, counting and recalling objects, or general situational awareness.
The widely used Mini Mental State Examination (MMSE), the Clock Drawing Test (CDT), the Mini-Cog test, the German SKT \cite{erzigkeit15}, among other cognitive scales have gained acceptance since they are brief while still showing good sensitivity and specificity \cite{sheehan12}.
Neuropsychological test batteries such as the Boston Diagnostic Aphasia Exam (BDAE) \cite{borod80}) and the CERAD-NB \cite{morris89} evaluate various perceptual modalities (auditory, visual, gestural), processing functions (comprehension, analysis, problem-solving), and response modalities (writing, articulation, manipulation).
They include common sub-tests such as the Cookie Theft Picture Test (CTP), the Boston Naming Test (BNT), and the Verbal Fluency Test (VFT).
Additionally, history taking interviews assist in looking for further dementia indicators related to activities of daily living (ADL), mood, physical activity and more.
Such interviews and tests are administered by trained physicians or psychologists who spend about 30--60 minutes with the patient.
With waiting times for appointments frequently exceeding six months, automated dementia screening could help to monitor patients closely and prioritize urgent cases for in-person appointments.

The automation of dementia screening based on speech is an area of high interest; it was previously covered by the ADReSS and ADReSSo challenges \cite{adress20,adresso21}.
Previous work shows strong evidence for the effectiveness of speech-based screening in dementia patients, even at early stages, and focuses primarily on the publicly available DementiaBank \cite{becker1994dementiabank}.
\cite{adress20,fraser16,alhameed16,koenig15,orimaye17} obtained convincing results on spontaneous speech of the CTP from the BDAE.
Free recall tasks of visual material, such as the CTP, have the advantage of eliciting speech on a common topic, making it more self-contained and thus easier to process.
The same is true for elicited speech based on free recall tasks from moving images, such as short films \cite{vincze22}.
Most work uses either fluency \cite{koenig18,frankenberg21} or deep speech markers \cite{adresso21} for classification, as these show high selectivity for discriminating patients with cognitive impairment from healthy controls.

This paper describes and reports the results of baseline experiments on the automated evaluation of a semi-structured clinical interview and three standardized sub-tests from the Syndrom-Kurz-Test (SKT, translates to ``Syndrome Short Test'') and the Consortium to Establish a Register for Alzheimer's Disease Neuropsychological Battery (CERAD-NB).
The speech data used in the experiments comprise 101 recordings of conversations between patients and psychologists collected during dementia screening sessions at the Memory Clinic of the Department of Psychiatry and Psychotherapy, Nuremberg Hospital in Germany. 
In our experiments, Support Vector Machine (SVM) classifiers are used in conjunction with openSMILE (OS) and wav2vec 2.0 (W2V2) features to test the feasibility of using speech data to automatically evaluate three sub-tests and categorize patients into cognitively impaired and non-impaired.
In addition, we investigate whether this classification is possible using short segments of spontaneous speech extracted from the clinical interview.

\section{Data}\label{sc:data}
All dementia screenings were carried out at the Memory Clinic (``Ge\-dächt\-nis\-sprech\-stunde'') of the Department of Psychiatry and Psychotherapy, Nuremberg Hospital, Germany.\footnote{Research approved by the Ethics Committee of the Nuremberg Hospital under File No. IRB-2021-021; each subject gave informed consent prior to recording.}
From an ongoing recording effort, to date, a total of 101 recordings of German-speaking subjects aged 55 to 88 years ($\mu = 73.9 \pm $8.5) have been acquired (40 male, 61 female).
Their medical diagnoses range from no or mild cognitive impairment to mild and moderate dementia.
The fact that the data includes patients with no cognitive impairment despite being referred to the Memory Clinic makes this data set somewhat unique: typically, such ``healthy controls'' would be recruited separately.

All participants underwent a three-part screening procedure: clinical interview (cf. \Cref{sec:interview}); SKT and CERAD-NB tests; two questionnaires for self-assessment of mood (GDS-K: Geriatric Depression Scale Short Form) and activities of daily living (B-ADL: Bayer-Activities of Daily Living Scale).

Data includes labels for SKT and CERAD-NB sub- and total scores, both as raw and normalized values, as well as coded medical and psychological diagnoses (work in progress).
Metadata includes sex, age, smoker/non-smoker, medication (antidementives, antidepressants, analgesics), GDS-K, B-ADL (self and informant assessment), NPI (Neuropsychiatric Inventory, informant assessment), IQ-range (below average, $<$90; average, 90--110; above average, $>$110), and years of education.
Furthermore, we labeled the data with start and end times for each of the sub-tests.

The audio recordings consist of 83.3 hours of speech and were performed with a Zoom H2n Handy Recorder in XY stereophonic mode, positioned between the patient and the psychologist in such a way that level differences between the left (psychologist) and right (patient) channels could be used to separate the speakers.
The audio samples were recorded in 16-bit stereo wav format with a sampling rate of 48 kHz and later converted to uncompressed PCM mono wav format with a sampling rate of 16 kHz.
Both psychologists and patients reported that they were not affected by the presence of the device.
Due to the Corona pandemic, psychologists and patients wore surgical or KN95 masks that affect the speech signal according to \cite{nguyen21}.
The speech of some subjects exhibits strong forms of local accents and dialects. 

\subsection{Standardized Sub-Tests}\label{sec:subtests}
The dataset contains recordings from screening sessions, including time-labeled segments with speech from subjects performing the standardized sub-tests of SKT and CERAD-NB.
We selected recordings from three sub-tests that we considered particularly suitable for classification experiments using only acoustic features; the time limit of all three tasks is one minute. 
The following section provides a brief description of the sub-tests and the data. 

\begin{figure}
\includegraphics[width=0.8\textwidth]{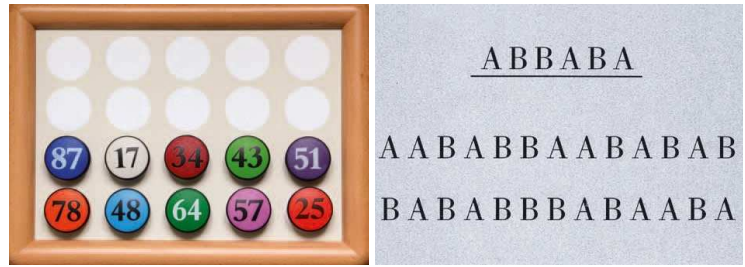}
\centering
\caption{Example templates of the sub-tests SKT 3 (left) and SKT 7 (right) from the original SKT Manual \cite{erzigkeit15}.}
\label{fig:skt3skt7}
\end{figure}

\subsubsection{SKT 3 (reading numbers)}

Sub-test SKT 3 starts with the psychologist asking the patient to read the two-digit numbers written on the colored game pieces (\figurename~\ref{fig:skt3skt7}, left) out loud in the direction of reading; this should be done as quickly as possible. 
We chose this reading task because of its simplicity compared to the other sub-tests; patients with mild impairments usually still perform well in it.
The time needed to complete the task is converted into norm values from 0 (no impairment) to 3 (severe impairment).
These values are normalized according to age and IQ-group (below average, average, above average) \cite{erzigkeit15}.
With a cut-off value of 1 (mild impairment), we separate the subjects into impaired (1--3) and non-impaired (0) and observe an almost balanced class distribution of 47/54 respectively.

\subsubsection{SKT 7 (interference test)}
Sub-test 7 is an interference test (\figurename~\ref{fig:skt3skt7}, right). It measures the ``disposition rigidity'' according to R.B. Cattell \cite{cattell49}, i.e., the mental ability to switch.
The aim is to learn to quickly break through intensively learned responses (here: the alphabet).
A sequence consisting of two repeating letters (e.g., ``A'' and ``B'') is to be read as quickly and accurately as possible.
The particular challenge is that the subject has to read one letter but say the other (i.e., read ``A'' but say ``B'' and vice versa).
The underlined letters serve to explain the task and are not to be worked on by the patient and thus are not included in the temporal evaluation.
We chose this interference test because it is comparatively the most demanding in terms of subjects' cognitive performance. 
It happens that more severely impaired patients do not understand the task or achieve only very low performance. 
The merit of this task lies in its sensitivity to mental performance impairment: Especially in the range of questionable or very mild impairments, it can differentiate best.
As in sub-test 3, the time required is converted into norm values from 0 to 3 and a cutoff value of 1 is set, resulting in a balanced class distribution of 50/51 for non-impaired and impaired subjects, respectively.

\subsubsection{CERAD-NB 1 (verbal fluency test)}
The CERAD-NB 1 is used to examine the speed and ease of verbal production ability, semantic memory, linguistic ability, executive functions, and cognitive flexibility. 
The psychologist conducting the sub-test asks the patient to name as many different animals as possible within one minute; the number of correctly named animals forms the basis for the test score.
We choose this verbal fluency test (VFT) because it has already been shown to be suitable for our purpose in related work \cite{koenig18,frankenberg21}. 
The CERAD-NB 1 raw values (number of named animals) are normalized taking into account the influence of age, education level, and sex according to \cite{berres00} and then converted to z-scores.
The z-score indicates by how many standard deviations a found value deviates from the mean of the population.
Statistics of the studied healthy norm population from \cite{aebi02} are used as reference. 
However, there are inherent selection biases in the overall (Memory Clinic) and study (mildly impaired) populations, and at the time of writing there is no compensating healthy control group, making class separation considerably more difficult.
While SKT 3 and SKT 7 address the patient's mental attention, CEARD-NB 1 differs in execution and examines mental production.
To obtain a conclusive class division despite these limitations and differences, we calculate the z-score threshold for CERAD-NB 1 based on the individuals matched in the classes for SKT 3 and SKT 7 (73\%), i.e., between the two groups for true-positive (impaired) and true-negative (non-impaired).
The resulting z-score of $-1.2$ leads to a balanced distribution of 50/51 for the non-impaired ($>-1.2$) and impaired ($\leq-1.2$) classes in our data set.


\subsection{Clinical Interview}\label{sec:interview}
The semi-structured clinical interview includes questions on memory, orientation, attention, activities of daily living, mood, sleep quality, appetite, physical activity, and medication of the patient.
It also includes an intro (greeting and introduction of the interview part) at the beginning and a final part (introduction of the testing part) by the psychologist.
For this reason, we extracted samples (4$\times$30 sec) from the middle (at 30\%, 40\%, 50\% and 60\%) of the interview to capture as much patient speech as possible; ground truth diarization was available from manual transcriptions of 30 patients.

For the interview samples, we use the CERAD-NB 1 labels as targets since speech of the VFT is inherently more similar to the spontaneous speech from the interview and also allows for more deep speech markers than the other two tasks (more in \cref{sc:discussion}).

\section{Methods}
This section briefly describes the features used for the machine learning experiments conducted. 
\subsection{openSMILE}
OpenSMILE is a popular toolkit that is used for the extraction of audio features \cite{eyben_OpensmileMunichVersatile_2010}.
The toolkit computes several functionals over low-level descriptor (LLD) contours in a brute-force approach. 
In our experiments, we use the ComParE 2016 feature set, which consists of 6373 static features.
OpenSMILE features are widely used in baseline experiments.
The features have been shown to achieve proper baseline performance in numerous paralinguistic applications such as gender detection, age detection, or speech emotion recognition  \cite{schuller_interspeech_2016,schuller_interspeech_2021}.

\subsection{wav2vec 2.0}
Models based on transformer architectures achieve state-of-the-art performance in various areas. 
Early breakthrough results have been achieved in the natural language processing domain \cite{devlin19}.  
Wav2vec 2.0 (W2V2) is a neural network model based on the transformer architecture designed for learning speech representations from raw audio data. 
The model is usually pre-trained with large amounts of unlabeled audio data \cite{baevski_wav2vec_2020}.
The W2V2 model takes raw waveform audio data as inputs, which are processed by a Convolutional Neural Network (CNN) encoder, followed by a contextualized transformer network and a quantization module.
The CNN generates latent representations directly from the waveform inputs, which are then discretized by the quantization module.
The convolutional feature extractor is followed by twelve contextualized transformer blocks that use self-attention to make the model focus on the parts of the audio relevant to respective tasks.
The model can be used as a feature extractor with or without adaptation.

W2V2 features are contextualized audio representations that encode information about a vector's relationship at time step $t$ to other vectors inside the extraction window \cite{baevski_wav2vec_2020}.
Due to the way transformer models are trained, they are capable of extracting many aspects of the underlying data.
The W2V2 model yields different speech representations after every transformer block, encoding the audio depending on the position in the processing hierarchy.
Thus, the representations after each layer focus on different aspects of the encoded audio, making them more or less useful for a particular task \cite{baevski21}. 
Features extracted from the model have been successfully applied to tasks such as phoneme recognition, speech emotion detection, and mispronunciation detection \cite{baevski_wav2vec_2020,pepino_emotion_2021,xu_explore_2021}.

The W2V2 features used in our experiments were extracted from models pre-trained unsupervised on 960 hours of speech from the LibriSpeech corpus.
We hypothesize that features extracted using only the weights obtained with unsupervised training will emphasize fine-grained differences in speech, as opposed to features that were fine-tuned for speech recognition, since these must reduce differences to be more robust w.r.t. articulation and speech production in order to increase robustness of speech recognition.
The model takes the raw waveform audio data as inputs and returns 768-dimensional speech representations after each transformer block, representing roughly 0.02 seconds of audio.
This yields $N = T/0.02 - 1$ vectors for the extraction context of T, i.e. 449 vectors with the extraction context of 10 seconds used.
For the speech data of each sub-test, we extract features and, analogous to mean pooling along the time dimension, compute a mean vector over all extracted feature vectors of a sample. As a result, we obtain one vector representing the audio of the respective sub-test. 
For the interview audio data, we take the mean for the samples extracted at the specified relative duration and perform the same processing, yielding four vectors for each subject.

\section{Experiments}

Our experiments aim to differentiate the speech of individuals who are cognitively impaired from the speech of individuals who are not cognitively impaired in the context of their performance on neuropsychological tests.
The experiments are conducted with speech data from the three sub-tests described in \cref{sec:subtests} and with speech data extracted at specific points in the semi-structured interview described in \cref{sec:interview}, relative to the duration of the interview.

The experiments using speech data from the standardized sub-tests are similar to experiments conducted with data from the ADReSSo challenge, which includes recordings of patients and healthy control speakers performing the CTP.
Since subjects are asked to perform a standardized task in a given time, the speech samples should be inherently comparable, making them ideal for experimentation. 
There are however some limitations to our experiments:
At the time of writing, we do not have an independent healthy control group, which is why we relate the labels for cognitive impairment to performance on the three sub-tests. 
Thus, performance on the sub-tests is not necessarily equivalent to the subject's diagnostic cognitive state.
We choose labels for impaired and non-impaired for SKT 3, SKT 7, CERAD-NB 1 according to \cref{sec:subtests} and for the interview speech samples according to \cref{sec:interview}.

We use Support Vector Machine (SVM) classifiers with radial basis function kernels (rbf) as they allow for quick experiment turnaround and are able to learn from only a few samples.
The optimal hyperparameters for the SVM and the respective input features for the SVM classifiers were determined using grid search in stratified five-fold cross-validation on the respective training portion of the data.

We use five-fold cross-validation of disjoint speakers.
For the sub-tests, the data was split into five distinct training sets comprising ~80\% of the data and test sets comprising the remaining ~20\%.
The training test partitioning of the interview segment data (4 segments/speaker) uses stratified group partitioning for speaker-exclusive folds considering label distribution.
Each training portion is then split again into five folds to determine the best hyperparameters in a stratified five-fold cross-validation.
The kernel parameter $\gamma$ was selected from the set
$ \gamma \in \{10^{-k} \mid k = 1, \ldots, 5 \} \subset \mathbb{R}_{>0} $,
the regularization parameter $C$ was selected from
$C \in \{10^{k} \mid k = -1, \ldots, 3 \} \subset \mathbb{N} $, and specific to the experiments conducted using W2V2 features, the W2V2 extraction layer $L$ was selected from $L \in \{1, 2, \ldots, 12\}$.

We evaluate our models' performance by accuracy, which is a good indicator of model performance since the data set is mostly balanced between the classes.

\subsection{Results}

\begin{table}[t]
    \centering
    \caption{Average classification accuracy and standard deviation (in \%) over the five test folds using OpenSMILE (OS) and wav2vec 2.0 (W2V2) features for SKT 3, SKT 7, CERAD-NB 1, and the interview (predicted on CERAD-NB 1 label). For W2V2, the best numbers after investigating the classification performance of features taken from the 12 different layers of the model are shown.}
    \label{tab:results}
    \begin{tabular}{c||c|c|c|c}
\toprule
\textbf{~~Method~~} & \textbf{~~SKT 3~~} &  \textbf{~~SKT 7~~}  &  \textbf{~~CERAD-NB 1~~} & \textbf{~~Interview~~}\\
\midrule
OS & $78.1\pm5.4$ & $84.8\pm2.1$ & $67.6\pm7.1$ & $53.5\pm3.3$\\
W2V2 & $82.9\pm4.3$ & $84.8\pm7.1$ & $77.1\pm8.5$ & $67.3\pm4.4$\\
\bottomrule
\end{tabular}
\end{table}

\tablename~\ref{tab:results} contains the experimental results. 
We report the average classification accuracy over all five test folds using OS and W2V2 features.

With OS features, we observe solid classification results on the sub-tests SKT 3, SKT 7 and CERAD-NB 1 with accuracies of 78.1\%, 84.8\% and 67.6\%, respectively.
Using the speech samples taken from the interview, OS features do not seem to provide any discriminatory power, leading to results at chance level. 

\begin{figure}[t]
\includegraphics[width=0.85\textwidth]{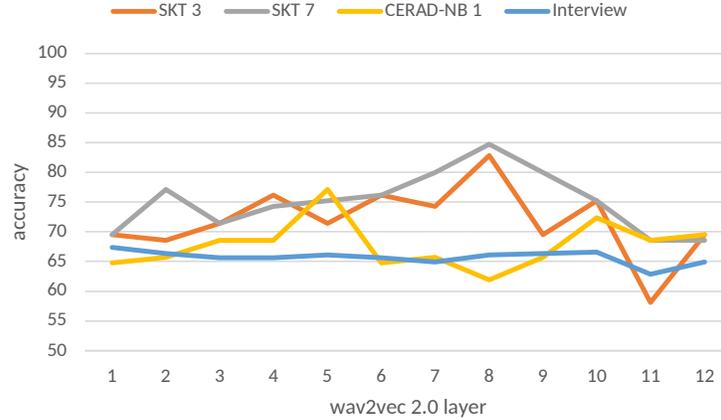}
\centering
\caption{Average classification accuracy over the five test folds for features taken from each of the 12 wav2vec 2.0 layers for SKT 3, SKT 7, CERAD-NB 1, and the interview.} \label{fig:w2v2_acc}
\end{figure}

\tablename~\ref{tab:results} contains the best results using W2V2 features.
We investigated the classification performance w.r.t. the features taken from the 12 different layers of the W2V2 model.
To that end, \figurename~\ref{fig:w2v2_acc} depicts the performance of classifiers utilizing W2V2 features taken from each of the 12 layers for each task. 
Experiments on the constrained tasks of reading numbers (SKT 3) and the interference test (SKT 7) achieve adequate to high accuracies on all W2V2 layers, with both reaching their maximum accuracy of 82.9\% (SKT 3) and 84.8\% (SKT 7) on layer 8.
This suggests that markers of cognitive status for constrained speech tasks are amplified in the upper-middle layers of the processing hierarchy.
The VFT is the task of naming animals (CERAD-NB 1), which is intrinsically more open-ended than the other two tasks focusing on production instead of attention; the speech and content will therefore vary more from patient to patient.
Nevertheless, experiments on CERAD-NB 1 are promising, yielding an accuracy of up to 77.1\% using features extracted from W2V2 layer 5. 
For the selected speech segments from the interview, the average accuracy does not vary much across the layers, ranging around 66\%.
Here, we obtain the best classification result on layer 1 with 67.3\% accuracy.
We therefore hypothesize that spontaneous speech taken from a semi-structured interview may be sufficient to extract discriminating speech features that can help with the detection of cognitive impairment.

\section{Discussion}\label{sc:discussion}
Even though we achieve partly high accuracies on the SKT 3 and SKT 7, it is important to question whether the features represent ``deep speech markers'' that lead to these results or whether they rather capture basic features such as delays and rate of speech.
It is noticeable that OS features perform as well as W2V2 features on the SKT 7. 
This could be due to the fact that this sub-test has a high sensitivity for mental performance impairment, which in turn is reflected in basic acoustic features, such as the ones extracted with OS.
It is becoming clear that there will be no one-fits-all method for automating the entire SKT and CERAD-NB test inventories. 
This may be well suited for the CTP, as it captures both attention (timing constraints) and production (picture description) in one test and thus allows screening for dementia in general.
However, we focus on test inventories that intentionally cover a number of different neuropsychological domains with specific tests in order to obtain a detailed diagnostic picture of the patient, which therefore will also require a differentiated investigation in methodology.
Thus, an important finding for us is the question of which sub-tests are actually suitable for acoustic evaluation and which sub-tests should rather be evaluated at a textual or even semantic level. 
All the more we would like to emphasize the result that the classification on the spontaneous speech of the interview already worked in our basic experiment with random samples.
Manual transcriptions for all patients, which are in progress, will allow the targeted selection of patient speech in the interview and thus a more accurate interpretation of the results.
Once the medical and psychological diagnoses are finalized, a detailed analysis of the diagnoses, e.g., Alzheimer's dementia (AD) or mild cognitive impairment (MCI), of the misclassified individuals could be helpful in understanding and improving the results.
For example, AD patients are presumably more likely to be identified in language production, whereas MCI patients should be evaluated semantically.

\section{Conclusion}
We successfully classified cognitive impairments in three neuropsychological sub-tests, namely the SKT 3, SKT 7, and CERAD-NB 1, by using OS and W2V2 features from the elicited speech of these standardized tests to train SVM classifiers.
Using OS features, we demonstrated high accuracies of 78.1\% (SKT 3), 84.8\% (SKT 7), and 67.6\% (CERAD-NB 1), which remained the same for SKT 7 but improved to 82.9\% and 77.1\% for SKT 3 and CERAD-NB 1, respectively, when using W2V2 features.
We found that constrained speech (SKT 3 and 7) performed best at level 8, while speech from a fluency task (CERAD-NB 1) performed best at level 5.
Spontaneous speech (interview), on the other hand, showed similar performance on all layers, with layer 1 performing slightly better than the others.
In addition, we provided conclusive evidence that spontaneous speech from the interview can be used to extract discriminating features for the detection of cognitive impairment.

The task of automating test inventories such as the SKT and the CERAD-NB is difficult, and there will probably never be just one universal method to accomplish this.  
Just as the original tests are an ensemble of specialized sub-tests that target different neuropsychological domains, tailored methods will be needed to automatically evaluate the sub-tests. 
In the future, the analysis of completed diagnoses and the inclusion of a healthy control group will help to define more distinct classes. 
For the experiments on the spontaneous speech of the interview, automatic as well as manual transcriptions including speaker diarization will help to target patient speech only and factor out the potential influence of the interviewer \cite{perez-toro_influence_2021}.


%


\bibliographystyle{splncs04}
\bibliography{tsd1158a,tsd1158b}
\end{document}